\documentclass[12pt]{article}

\usepackage{geometry}
\usepackage{lscape}
\usepackage{array}

\usepackage{graphicx}
\usepackage{amsmath}
\usepackage{bbm}
\usepackage{cite}
\usepackage[sort&compress,numbers]{natbib}
\usepackage{amsfonts}
\usepackage{amssymb}
\usepackage{amsmath}
\usepackage{latexsym}
\usepackage{xcolor}
\usepackage{url}
\usepackage{ntheorem}
\usepackage{braket}
\usepackage[normalem]{ulem}
\usepackage{enumerate}
\usepackage[colorlinks=true]{hyperref}

\newcommand{\whhyp}[1]{\,\,\,\blue{\widehat{\!\!\!{\hyp_{\, #1}}}}}
\newcommand{\zhyp}{\,\, \blue{\mathring{\!\!\hyp}}}

\newcommand{\magenta}[1]{{\textcolor{magenta}{#1}}}

\newcommand{\ptcheck}[1]{\ptc{checked on #1}}

\newcommand{\mcO}{\mycal O}

\newcommand{\redM}{\red{M}}

\newcommand{\fourg}{\red{\mathfrak{g}}}

\DeclareFontFamily{OT1}{rsfs}{}
\DeclareFontShape{OT1}{rsfs}{CGNPm}{n}{ <-7> rsfs5 <7-10> rsfs7 <10-> rsfs10}{}
\DeclareMathAlphabet{\mycal}{OT1}{rsfs}{CGNPm}{n}

{\catcode `\@=11 \global\let\AddToReset=\@addtoreset}
\AddToReset{equation}{section}

{\catcode `\@=11 \global\let\AddToReset=\@addtoreset}
\AddToReset{figure}{section}

{\catcode `\@=11 \global\let\AddToReset=\@addtoreset}
\AddToReset{table}{section}

\newcommand{\R}{\mathbb R}

\newcommand{\hyp}{ \mycal S}

\newcommand{\blue}[1]{{\color{blue}#1}}

\newcommand{\mcD}{{\mycal D}}

\newcommand{\mcL}{\blue{\mycal L}}

\newcommand{\bea}{\begin{eqnarray}}
\newcommand{\beaa}{\begin{eqnarray*}}
\newcommand{\bean}{\begin{eqnarray}\nonumber}

\newcommand{\bel}[1]{\begin{equation}\label{#1}}
\newcommand{\beal}[1]{\begin{eqnarray}\label{#1}}
\newcommand{\beadl}[1]{\begin{deqarr}\label{#1}}
\newcommand{\eeadl}[1]{\arrlabel{#1}\end{deqarr}}
\newcommand{\eeal}[1]{\label{#1}\end{eqnarray}}
\newcommand{\eead}[1]{\end{deqarr}}
\newcommand{\eea}{\end{eqnarray}}
\newcommand{\eeaa}{\end{eqnarray*}}

\newcommand{\Ricc}{\mathrm{Ric}\,}

\newcommand{\be}{\begin{equation}}
\newcommand{\ee}{\end{equation}}

\newcommand{\tr}{\mbox{\rm tr}\,}

\newcommand{\myphi}{\red{N}}

\newcommand{\riemgz}{g_0}

\renewcommand{\hbar}{{\overline \riemgz}}

\newcommand{\red}[1]{{\color{red}#1}}

\newcounter{mnotecount}[section]

\renewcommand{\themnotecount}{\thesection.\arabic{mnotecount}}

\newcommand{\mnote}[1]
{\protect{\stepcounter{mnotecount}}$^{\mbox{\footnotesize
$
\bullet$\themnotecount}}$ \marginpar{
\raggedright\tiny\em
$\!\!\!\!\!\!\,\bullet$\themnotecount: #1} }

\newcommand{\ptc}[1]{\mnote{{\bf ptc:}#1}}

\newcommand{\plus}{+}
\newcommand{\minus}{-}

\renewcommand{\blue}[1]{#1}
\renewcommand{\red}[1]{#1}
\renewcommand{\magenta}[1]{{{#1}}}
\renewcommand{\ptcheck}[1]{}

\begin{document}
\title{Analyticity of stationary spacetimes from maximal hypersurfaces\thanks{Vienna preprint  UWThPh 2022-9}
}
\author{
Piotr T. Chru\'sciel\thanks{Email \protect\url{piotr.chrusciel@univie.ac.at}, URL \protect\url{http://homepage.univie.ac.at/piotr.chrusciel/}}
\\
Faculty of Physics,
University of Vienna
\\
\phantom{x}
\\
Marc  Mars\thanks{Email \protect\url{marc@usal.es}}
\\
Faculty of Sciences, University of Salamanca}
\date{}

\maketitle


\begin{abstract}
The standard method of proving analyticity of stationary vacuum metrics invokes the quotient-space version of Einstein equations. We verify that the same conclusion can be obtained using the KID equations on maximal surfaces.
\end{abstract}

\tableofcontents

\section{Introduction}

A classical theorem of Morrey~\cite{MorreyInterior} asserts that solutions of a class of elliptic PDEs are real-analytic. We verify that this theorem applies to vacuum Lorentzian metrics with a Killing vector field which is transverse to a maximal surface if and only if the Killing vector is timelike. 
This provides another example of usefulness of maximal hypersurfaces in general relativity. 

For definiteness we restrict attention to the vacuum Einstein equations with  cosmological constant, but the argument applies as is to all matter models where the field equations do not involve second derivatives of the metric and become elliptic when stationarity is assumed.

We note that from the point of view of general relativistic applications, e.g.\ to black hole uniqueness theorems,  the usefulness of this result is unique continuation of solutions, which follows from analyticity,  even though analyticity may not be necessary in some cases to obtain unique continuation; 
see \cite{ChDelayUnique} and references therein.

\section{ADM notation and Killing vector fields}

Any Lorentzian metric $\fourg$ can be locally written in the Arnowitt-Deser-Misner form
(cf., e.g., \cite[Equation~(21.40), p.~507]{MTW})
\begin{equation}\label{23VI22.1}
  \fourg= - N^2 dt^2 + g_{ij}(dx^i + Y^i dt)(dx^j + Y^j dt)
  \,,
\end{equation}
where $g_{ij}dx^i dx^j$ is a family of possibly $t$-dependent Riemannian metrics on a manifold $M$, $N$ is a function on spacetime and $Y=Y^i\partial_i$ is a family of possibly $t$-dependent vector fields on $M$.

Let $T$ be the field of unit normals to the level sets of $t$, one finds
\begin{equation}\label{23VI22.5}
  T = N^{-1}
   ( \partial_t - Y^i \partial_i )
  \qquad
  \Longleftrightarrow
  \qquad
  \partial_t = N T + Y
   \,.
\end{equation}
Since $Y$ is tangent to $\hyp$,  the vector $\partial_t$ is transverse to $\hyp$ if and only if $N$ has no zeros.

It follows from \eqref{23VI22.1} that
\begin{equation}\label{23VI22.2}
  \fourg_{tt} \equiv \fourg(\partial_t,\partial_t) = - N^2 + g_{ij} Y^i Y^j \equiv -N^2 +g(Y,Y)
  \,,
\end{equation}
which shows that
$\partial_t$ is spacelike if and only if the $g$-length of $Y$ is larger than $N$.

We have
\begin{equation}\label{23VI22.3}
  \det \fourg = - N^2 \det g
  \,,
\end{equation}
which shows that the metric \eqref{23VI22.1} is manifestly Lorentzian in the coordinate system above when  $N$ has no zeros and $g$ is Riemannian. This holds regardless of the $g$-length of $Y$.

The extrinsic curvature tensor (second fundamental form) of the level sets of $t$ is given by
\begin{equation}\label{23VI22.9}
  K_{ij} =
      \frac 12 \mcL_T g_{ij} =   \frac 12 N^{-1} ( \partial_t g_{ij} -  \mcL_Y g _{ij})
  \,,
\end{equation}
and we note that sometimes an opposite sign convention is used in \eqref{23VI22.9}.

If all the functions appearing in \eqref{23VI22.1} are $t$-independent the vector field $\partial_t$ is a Killing vector field:
\begin{equation}\label{23VI22.6}
  0 = \partial_t \fourg_{\mu\nu} = \mcL_{\partial_t} g _{\mu\nu}
  \,.
\end{equation}
In this case \eqref{23VI22.9} can be rewritten as
\begin{equation}\label{23VI22.8}
   \mcL_Y g _{ij}  = \minus 2 N K_{ij}
   \qquad
   \Longleftrightarrow
   \qquad
   D_{(i}Y_{j)} = \minus N K_{ij}
  \,,
\end{equation}
where $D $ is the Levi-Civita covariant derivative associate with the Riemannian metric $g=g_{ij}dx^idx^j$, and where parentheses over indices denote symmetrisation.

Given a Lorentzian metric $\fourg$ with a Killing vector field $X$ one can construct local coordinates sastisfying \eqref{23VI22.1} and the first equation in \eqref{23VI22.6}, this proceeds as follows: Let $\hyp$ be any spacelike hypersurface which is  transverse to $X$. Let $t$ be a function defined near $\hyp$ by solving the problem
\begin{equation}
\label{23VI22.31}
  \mcL_X t=1\,, \qquad  t|_\hyp = 0
  \,.
\end{equation}
Let $y^i$ be any local coordinates on a coordinate patch $\mcO\subset \hyp$, we can define functions $x^i$ near $\mcO$ by solving the equations
\begin{equation}
\label{23VI22.33}
  \mcL_X x^i=0\,, \qquad  x^i|_\hyp = y^i
  \,.
\end{equation}
One checks that in these coordinates the metric can be written in the form \eqref{23VI22.1}. We have $X^t = X(t)= \mcL_X t=1$ and $X^i = X(x^i)= \mcL_X x^i=0$, which shows  that $X=\partial_t$. Transversality of $X$ to $\mcO$ is equivalent to the condition that $N$ has no zeros on $\mcO$.

Since the problem addressed is purely local, for our purposes here we can without loss of generality assume that $\mcO=\hyp$, and that the coordinate system above is global on $I\times \hyp$, where $I\subset \R$ is an interval containing $0$.

\section{Einstein equations}
 When $\partial_t$ is a Killing vector, so that $\partial_t g_{ij}=0= \partial_t K_{ij}$, the vacuum Einstein equations for the metric \eqref{23VI22.1} imply the following set of equations~\cite{MTW,ChDelayExotic,MaertenKIDs,ChBeigKIDs,Moncrief75}:%
 \footnote{The explicit form of these equations differs across references by conventions on signs; we follow \cite{ChBeigKIDs}.
 Our equations \eqref{28I22.1}-\eqref{28I22.2} coincide  with those of \cite{MaertenKIDs}, with our $Y$ denoted by $X$ there, after taking into account the constraint equations \eqref{30I22.31-}-\eqref{30I22.31}.
 }
%
%
%
\bea
 D _i D _j \myphi
\nonumber
 &= &
   \big(
   \Ricc(g)_{ij} - \red{\frac{2\Lambda}{n-2} g_{ij}}
    - 2K^{\red{\ell}}{}_iK_{\red{j\ell }} +\tr_gK K_{ij}
 \big)
        \myphi
\\
&&
  \minus  D _{\red{\ell}} K_{ij}   Y^{\red{\ell}}
   \minus  2K^{\red{\ell}}{}_{(i}D _{j)}Y_{\red{\ell}}
 \,,
  \label{28I22.1}
\\
 D _{(i}Y_{j)}
  & = & \minus K_{ij}\myphi
 \,.
  \label{28I22.2}
\eea
Here 
$\Ricc(g)_{ij}$ the Ricci tensor of $g$ and $R$ its trace, $\Lambda$ the cosmological constant and $n$ is the spacetime  dimension.
These equations  should be complemented by the vacuum constraint equations~\cite{BartnikIsenberg,CarlottoLR},
\begin{align}
      R
    +(\tr_gK )^2
    -  K^{ij}K_{ij} & = \red{2 \Lambda}
     \,,
      \label{30I22.31-}
\\
    D ^{p}K_{\red{\ell} p}- D _{\red{\ell}}  \tr_gK
     & = 0
    \,.  \label{30I22.31}
\end{align}

Summarising: Let the $t$-independent fields $(\myphi,Y, g)$ parameterise the spacetime metric $\fourg$ as
\begin{equation}\label{19XII21.12}
  \fourg= -\myphi^2 dt^2 + g_{ij}(dx^i + Y^i dt)(dx^j + Y^j dt)
  \,.
\end{equation}
Then the metric $\fourg$ has a  Killing vector $X=\partial_t$ and satisfies the vacuum Einstein
equations if and only if \eqref{28I22.1}-\eqref{30I22.31} hold. The function $N$ will have no zeros on $\hyp=\{t=0\}$ if the Killing vector is transverse to $\hyp$.

In the situation just described we can use \eqref{28I22.2} to write $K_{ij}$ as
\begin{align*}
K_{ij}= - \frac{1}{2\myphi} \left ( D_i Y_j + D_j Y_i  \right )
\end{align*}
and insert this  into equations \eqref{28I22.1} and \eqref{30I22.31}. A direct computation gives
\begin{align}
\Ricc(g)_{ij}  = &
\frac{1}{\myphi} \left ( D_i D_j \myphi
- Y^{\ell} D_{\ell} \left ( \myphi^{-1} D_{(i} Y_{j)}  \right )
 + \frac 1{2 N} (D^\ell Y_i D_\ell Y_j - D_i Y^\ell D_j Y_\ell)
\right ) \nonumber \\
& - \frac{D^{\ell} Y_{\ell}}{\myphi^2} D_{(i} Y_{j)} \red{+ \frac{2\Lambda}{n-2} g_{ij}}
, \label{Ricg} \\
D^i D_{(i} Y_{j)}
= &   \myphi D_j \left ( \frac{D_i Y^i}{\myphi} \right )+
  D_{(i} Y_{j)}  D^i \ln |\myphi|. \label{LapY}
\end{align}
The trace of \eqref{Ricg} gives, after using the constraint \eqref{30I22.31-},
\begin{align}
\Delta_g \myphi =  \frac{1}{ \myphi}  D_i Y_j D^{(i} Y^{j)} +  Y^i D_i \left (\frac{D_j Y^j}{\myphi} \right ) \red{- \frac{2 \Lambda }{n-2} \myphi}
. \label{LapN}
\end{align}
%

\section{Maximal surfaces}

In the setting above, we claim that we can find in $\R\times \hyp$ a spacelike hypersurface, say  $\zhyp$,  on which it holds
\begin{equation}\label{23VI22.21}
  \tr_g K = 0
  \,,
\end{equation}
and which is transverse to the Killing vector $X$. (While eventually we will be interested in timelike Killing vectors, we do not make the assumption that $X$ is timelike here.)
While this should be clear to those which are familiar with the results of~\cite{Bartnik84}, in order to dispell any doubts we give a formal proof:
Let $p\in \hyp$ and consider a small coordinate ball $B(0,r)$ of radius $r$, centered at $p$, within $\hyp$. Let $\epsilon>0$ and let  $\phi$ be any smooth function on  $S(0,r)$ satisfying
\begin{equation}\label{23VI22.51}
 |\phi|+ |d \phi|_g<\epsilon
  \,.
\end{equation}
There exists $r$ small enough and $\epsilon$ small enough so that for all functions $\phi$ satisfying \eqref{23VI22.51} we can find a spacelike hypersurface $\whhyp{}$ such that:
\begin{enumerate}
  \item $\partial \whhyp{}:= \overline{\whhyp{}}\setminus \whhyp{} $ is the graph $\{t=\phi(p)\,, p\in \partial B(0,r)\}$;
  \item  the domain of dependence $\mcD(\whhyp{})$ of $\whhyp{}$ is globally hyperbolic;
  \item
and $\mcD(\whhyp{})$ has compact closure.
\end{enumerate}

By \cite[Theorem~4.2]{Bartnik84} there exists in $\R\times \hyp$ a  spacelike hypersurface, say $\hyp_\phi$ spanned on $\partial \whhyp{} $ and satisfying
\eqref{23VI22.21}.
Clearly $X$ cannot be tangent to all the hypersurfaces $\hyp_\phi$ obtained in this way. Hence
there exists a function $\phi$ such that $X$ will be transverse to $\hyp_\phi$ somewhere. We define  $  \zhyp$ to be any open subset of $\hyp_\phi$ to which $X$ is transverse.

Let us use again the symbol $g$ for the metric induced by $\fourg$ on
$\red{\zhyp}$.
We can carry-out the construction,  described around  \eqref{23VI22.31}-\eqref{23VI22.33}, of a new coordinate system, still denoted by $(t,x^i)$, by requiring that the new function $t$ vanishes on $\red{\zhyp}$. The metric $\fourg$ expressed in these new coordinates takes again the form \eqref{19XII21.12}, with $t$ equal to zero on $\red{\zhyp}$, with  new time-independent function $N$ and vector field $Y$ and metric $g$, and with a Killing vector $\partial_t$, Moreover \eqref{23VI22.21} holds.

The  vacuum KID equations \eqref{28I22.1}-\eqref{28I22.2} and the 
vacuum constraint equations \eqref{30I22.31-}-\eqref{30I22.31}
continue to hold on $\zhyp$.  Inserting $D_i Y^i =0$ into \eqref{Ricg}-\eqref{LapN} yields 
%
%
\begin{align}
&   \Delta_g N -    \frac{1}{\myphi}
D_i Y_j D^{(i} Y^{j)}   \red{+ \frac{2\Lambda}{n-2} N}
      =  0   \label{30V22.1-c}
\\
&   D^i D_{(i}Y_{j) } - D_{(i}Y_{j) }D^i \ln| \myphi|
    =  0   \label{30V22.1} \\
%
%
&  \Ricc(g)_{ij}
  -\frac{1}{N}
   \bigg(
   D _i D _j \myphi
  -   Y^{\red{\ell}} D _{\red{\ell}}( N^{-1}  D_{(i}Y_{j) })
   + \frac 1{2 N} (D^\ell Y_i D_\ell Y_j - D_i Y^\ell D_j Y_\ell)
  \bigg)    \nonumber \\
&  \hfill = \red{ \frac{2 \Lambda}{n-2} g_{ij}}
  \label{24VI12.3}
\end{align}

\section{Analyticity}

In harmonic coordinates the system \eqref{30V22.1-c}-\eqref{24VI12.3} forms an elliptic system for $(N,Y^i,g_{ij})$  if and only of $N^2 > |Y|^2_g$, as we explicitly show next.
In the notation of \cite{MorreyInterior}, the principal symbol $L(x,\lambda)$ of the system  \eqref{30V22.1-c}-\eqref{24VI12.3} is obtained by linearising the equations around the solution $(N,Y,g)$:
 for $\lambda\in T^*\zhyp $,
  \ptcheck{4VII22}
\begin{align}\label{24VI22.1}
  &L(x,\lambda)
  \left(
    \begin{array}{c}
      \delta N \\
      \delta Y \\
      \delta g \\
    \end{array}
  \right)
   = \\
   &  \left(
    \begin{array}{c}
       g^{k\ell }\lambda_k \lambda_\ell \delta N \\
      \frac 12 \lambda^i (\lambda_i \delta Y_j + \lambda_j \delta Y_i
\magenta{+\lambda^k Y_k \delta g_{ij}
- \lambda_i \delta g_{jk} Y^k
- \delta g_{ik} Y^k \lambda_j}
)
 \nonumber
\\
      -\frac 12 g^{k\ell }\lambda_k \lambda_\ell \delta g_{ij } - \frac{1}{N} \lambda_i \lambda_j \delta N
      + \frac{Y^{\ell}}{2 N^2}
       \Big(
          \lambda_\ell (\lambda_i \delta Y_j + \lambda_j \delta Y_i)
      + \magenta{Y^k \lambda_{\ell} \left (
      \lambda_k \delta g_{ij}
      - \lambda_i \delta g_{jk}
      - \lambda_j \delta g_{ik} \right )}
      \Big)
        \end{array}
  \right)
   \,.
\end{align}
Ellipticity for this system is defined in \cite{MorreyAnalyticity,FriedmanRegularity} as the condition that the determinant of the principal symbol is non-zero. This is equivalent to the requirement that the equation
\begin{equation}\label{24VI22.2}
  L(x,\lambda)
  \left(
    \begin{array}{c}
      \delta N \\
      \delta Y \\
      \delta g \\
    \end{array}
  \right)
   = 0
\end{equation}
has no non-trivial solutions unless $\lambda \equiv 0$. So suppose that $\lambda\ne 0$, then the equation
$$
g^{ij}\lambda_i \lambda_j \delta N =0
$$
gives $\delta N \equiv 0$, since $g$ is Riemannian.

Next, consider the equation
\begin{align}\label{24VI22.31}
       \lambda^i (\lambda_i \delta Y_j + \lambda_j \delta Y_i
\magenta{+\lambda^k Y_k \delta g_{ij}
- \lambda_i \delta g_{jk} Y^k
-  \lambda_j \delta g_{ik} Y^k
}
)=0  \,.
\end{align}
Contracting with $\lambda^j$ gives
$$
 0 
 =  2 |\lambda|^2_g \left (
  \, \lambda^j \delta Y_j - \magenta{\delta g_{jk} \lambda^j Y^k}
\right )  \magenta{ + \lambda^k Y_k \delta g_{ij} \lambda^i \lambda^j}
 \,,
$$
which for non-zero $\lambda$ \magenta{(which we assume from now on)} implies
\begin{equation}
\lambda^j \delta Y_j= \magenta{\delta g_{ij} \lambda^i Y^j -
\frac{\lambda^k Y_k}{2 |\lambda|^2_g} \delta g_{ij} \lambda^i \lambda^j}.
\label{deltaY_lambda}
\end{equation}
Inserting this into \eqref{24VI22.31} gives
\begin{align}
\magenta{\delta Y_j = Y^i\delta g_{ij}
+ \frac{\lambda^k Y_k}{|\lambda|^2_g} \left (
-\lambda^i  \delta g_{ij}
+ \frac{1}{2 |\lambda|^2_g}
\lambda_j \lambda^i \lambda^k \delta g_{ik}
\right )}
 \,. \label{deltaY}
\end{align}
which is compatible with \eqref{deltaY_lambda}.
Replacing $\delta N=0$ and \eqref{deltaY}  into the third line of \eqref{24VI22.2} gives, after a simple manipulation,
\begin{align*}
\frac{1}{2}
\left ( - |\lambda|^2_g
+ \frac{(\lambda^k Y_k)^2}{N^2} \right ) \delta g_{ij}
+ \frac{(\lambda^k Y_k)^2}{2 N^2 |\lambda|^2_g} \left (
- \lambda_i \lambda^k \delta g_{jk}
-\lambda_j \lambda^k \delta g_{ik}
+ \frac{1}{|\lambda|^2_g} \lambda_i \lambda_j \delta g_{kl} \lambda^k \lambda^l
\right )=0\,.
\end{align*}
Contracting with $\lambda^j$ one gets
\begin{equation}\label{7V22.1}
 \lambda^j \delta g_{ij} =0
  \,,
\end{equation}
which inserted back into the equation yields
\begin{align}
\left ( - |\lambda|^2_g
+ \frac{(\lambda^k Y_k)^2}{N^2} \right ) \delta g_{ij} =0\,.
\label{eq_deltag}
\end{align}
If $|Y|^2_g < N^2$ the term in parenthesis is negative for all $\lambda_i$,
so $\delta g_{ij}=0$ which inserted into \eqref{deltaY} gives $\delta Y_j=0$
and ellipticity is established. When $|Y|^2_g \geq N^2$, let $s$ be any vector orthogonal to $Y$ with norm
\begin{align*}
|s|^2_g = |Y|^2_g \left ( \frac{|Y|^2_g}{N^2} -1 \right ) > 0\,.
\end{align*}
Then $\lambda_j := Y_j + s_j$ is non-zero and the parenthesis in
\eqref{eq_deltag} is identically zero. Thus, for such $\lambda$  the only restriction on
$\delta g_{ij}$ is \eqref{7V22.1},
which leads to a kernel of    $
  L(x,\lambda)$ consisting of fields  $(\delta N, \delta Y_i,\delta g_{ij})  $ satisfying $\lambda^i \delta g_{ij} =0$,  $\delta N=0$  and $\delta Y_i = \delta g_{ij}Y^j$. We conclude that the system is \emph{not}
elliptic whenever $|Y|^2_g \geq N^2$.
\ptcheck{4VII22}

Real-analyticity of solutions \magenta{for $|Y|^2_g< N^2 $}
follows from \cite{MorreyInterior} by  choosing the  indices  $s_i$ and $r_j$ there as $s_i=1=r_j$.
Alternatively one can appeal to \cite{FriedmanRegularity}.%
\footnote{
Further references of interest in the context include~\cite{Blatt,MorreyBoundary,Cosner,MorreyNirenberg}.}

The result is sharp: for completeness we present families of elementary examples of non-analyticity for non-timelike Killing vectors in Appendices~\ref{App28VI22.1} and \ref{s24VI22.1}.
The reader might find some interest of its own in the reduction of the problem to a dynamical system in the setting of   Appendix~\ref{App28VI22.1}.
 
While maximality of the level sets of $t$ is not necessary for analyticity, a restriction on the level sets of $t$ with an elliptic flavour is certainly necessary: indeed, given an analytic spacetime we can always change the time slicing by changing $t$ to $t+f$, where $f$ is smooth but \emph{not analytic}, with $df$  sufficiently small so that the level sets of the new time function are again spacelike. The resulting new data $(N,Y,g)$ will not be analytic.

\appendix

\section{pp-waves}
 \label{App28VI22.1}

\def\RicSig{\mbox{Ric}^{\hyp   }}

The $pp$-waves provide examples of smooth non-analytic metrics with a null Killing vector.  The metric takes the form
\begin{align*}
  \fourg = 2 dv du + H(u,x,y) du^2 + dx^2 + dy^2.
\end{align*}
The vector field $X = \partial_v$ is null, covariantly constant and nowhere zero, hence transversal to any spacelike hypersurface. We chose the orientation so that $\partial_v$ is future. The metric $\fourg$ is
Ricci flat if and only if $H$ satisfies~\cite[Equation~(24.43), p.~384]{Exactsolutions}
\begin{align*}
 ( \partial_x^2  + \partial_y^2) H =0,
\end{align*}
i.e. it is harmonic in the $x,y$ coordinates. Moreover, $\fourg$ is locally flat if and only if $H$ is a polynomial of degree one in the coordinates
$x,y$~\cite[Equation~(24.43), p.~384]{Exactsolutions}.

It is trivial to construct
non-analytic solutions. An example to be used below is
$H = 1 + w(u) xy$, with $w(u) \in C^{\infty} (\mathbb{R})$ satisfying
$w(u) =0$ in $u \in (- \infty,0)$, and $w(u)>0$ for $u>0$. The metric is Minkowski in the domain $u  <0$ and has non-zero curvature for $u>0$, so there exists no coordinate system in which the metric is analytic near the point $u=x=y=0$.

Consider a hypersurface $\hyp   :=\{ v = f(u,x,y)\}$.
To compute the fields $(N,Y,g)$
on $\hyp$ we introduce adapted coordinates, as described in the main text. Let $t = v- f(u,x,y)$ and let us write $\fourg$ in the coordinate system $\{t,u,x,y\}$:
\begin{align}
  \fourg =  2 dt du
  + \left ( H + 2 \partial_u f \right ) du^2
  + 2 \partial_x f du dx + 2 \partial_x f du dy
  + dx^2 + dy^2. \label{PPmetric}
\end{align}
The hypersurface $\hyp$ is $\{ t=0\}$ so the induced metric
(the first fundamental form) is
\begin{align}
  g = \left ( H + 2 \partial_u f \right )  du^2 + 2 \partial_x f dudx + 2 \partial_y f du dy + dx^2 + dy^2. \label{dataPP1}
\end{align}
This is Riemannian (positive definite) provided that
\begin{align}
  F:= H + 2 \partial_u f  - (\partial_x f)^2 - (\partial_y f)^2>0 \label{defF},
  \end{align}
which we assume from now on. To compute $\myphi$ and $Y$ we compare
\eqref{PPmetric} with \eqref{23VI22.1}.
The absence of a $dt^2$ term gives
$\myphi^2 = g_{ij} Y^i Y^j$, which is simply a restatement of the fact that $X$ is null everywhere.  The cross-terms  $dt dx^i$ give $g_{ij} Y^j dx^i = du$, which can be solved for $Y$:
\begin{align}
  Y = \frac{1}{F} \left ( \partial_u
  - f_x \partial_x - f_y \partial_y \right ). \label{dataPP2}
\end{align}
The $g$-norm of this vector is  $g_{ij} Y^i Y^j = \frac{1}{F} = \myphi^2$. Selecting the unit normal
$T$ to be future directed requires $\myphi>0$ (cf. \eqref{23VI22.5}). Consequently,
\begin{align}
  \myphi = \frac{1}{\sqrt{F}}. \label{dataPP3}
\end{align}
One can now check by an explicit calculation that
\eqref{28I22.1}-\eqref{28I22.2} hold.

The metrics  \eqref{PPmetric} provide an interesting example where the maximal surface equation reduces to a polynomial dynamical system with a source.  Specifically, the trace of the
extrinsic curvature, as defined in \eqref{23VI22.9}, is given by
\begin{align*}
  \tr_g K  =  \minus \partial_u \myphi \plus \partial_x (\myphi \partial_x f) \plus
  \partial_y (\myphi \partial_y f).
\end{align*}
As already pointed-out we choose $H = 1+ w(u) x y$, with $w(u)$ as before,
and look for solutions of the maximal surface
equation $\tr_g K=0$ of the form
\begin{align*}
  f= f_0 (u) + f_1(u) (x^2 + y^2) + f_2(u) xy.
\end{align*}
The condition that  $\hyp   $ is maximal turns out to be equivalent
to the following system of ODEs, where a dot
denotes a derivative with respect to $u$:
\begin{align*}
  \ddot{f_0} & = -4 f_1 (1+2 \dot{f}_0 ), \\
  \ddot{f_1} & = 2 f_2 \dot{f_2} +  2 f_1 (4 f_1^2- f_2^2)+ \frac{1}{2} f_2 w \\
  \ddot{f_2} & = 2 f_2 \left ( 4 \dot{f}_1 + 4 f_1^2 - f_2^2 \right ) -2 f_1 w-\frac{1}{2} \dot{w}.
\end{align*}
For $u\leq 0$ we  take $f_0(u)= f_1(u)=f_2(u)=0$  and for $u \geq 0$ we take the unique solution of this
system with initial data $f_0(0)= f_1(0)= f_2(0)= \dot{f}_0(0) =
\dot{f}_1(0) = \dot{f}_2(0)=0$. Since $w(u)$ is $C^{\infty}(\mathbb{R})$ but not analytic, the same holds for $f_i(u)$, $i=0,1,2$. It is immediate that
the condition \eqref{defF} (i.e., $\hyp   $
being spacelike)  is satisfied in a
neighbourhood of the point $u=x=y=0$. This provides an essentially explicit example
of a maximal hypersurface with transversal null Killing vector field and non-analytic metric.

\section{Einstein-Rosen waves}
 \label{s24VI22.1}

Consider the collection of metrics of the form
\begin{equation}\label{24VI22.41a}
  \fourg = e^{2(\gamma-\psi)}(-dt^2 + dr^2) + e^{-2\psi} r^2 d\varphi^2 + e^ {2\psi} dz^2
  \,,
\end{equation}
where the functions $\psi$ and $\gamma$ depend only upon $t$ and $r$.   Taking $t
\in \R$, $z\in\R$, and viewing $(r,\varphi)$ as polar coordinates on $\R^2$,  we thus obtain a family of cylindrically symmetric metrics on $\R\times\R^3$. Regularity of the metric at the axis of rotation $\{r=0\}$ requires the vanishing of $\gamma$ there.

The metric \eqref{24VI22.41} will satisfy the vacuum Einstein equations (with vanishing cosmological constant) if and only if \cite{EinsteinRosen,Exactsolutions,AshtekarBicakSchmidtER}
\begin{align}
  \big(
   -\partial_t^2 + \partial_x^2 + \partial_y^2
   \big)
    \psi  & = 0
    \,,
        \label{24VI22.41}
\\
   \partial_t \gamma
 &   = 2 r
   \partial_t \psi
    \partial_r \psi
  \,,
    \label{24VI22.43}
\\
   \partial_r \gamma  & =    r \big(
   (\partial_t \psi)^2 +
   (\partial_r \psi )^2
   \big)
  \,.
    \label{24VI22.42}
\end{align}

The key equation is \eqref{24VI22.41}: Indeed, given any rotation-invariant solution of the wave equation $\psi$ on $\R\times\R^2$, we can integrate \eqref{24VI22.43}-\eqref{24VI22.42}, with $\gamma(0,0)=0$, to obtain $\gamma$.
We note that \eqref{24VI22.41} is the integrability condition for \eqref{24VI22.43}-\eqref{24VI22.42}.

The vanishing of $\gamma$ at  $\{r=0\,,\ t=0 \}$  is preserved in time by \eqref{24VI22.43}.
Smooth initial data lead, by evolution, to smooth spacetime metrics by general considerations. This is clear from \eqref{24VI22.41}-\eqref{24VI22.43} in any case except possibly at $r=0$, and can be directly verified at the axis of rotation by inspection of asymptotic expansions there.

Hence the set of smooth such metrics can be uniquely parameterised by the set of smooth rotation-invariant initial data for the $(2+1)$-dimensional wave equation in Minkowski space-time.
 
Consider a point $p\in \R^4$ at which we have
\begin{equation}\label{30VI22.1}(-(
\partial_t \psi)^2 + (\partial_r \psi)^2 - r^{-1} \partial_r \psi)|_p \neq 0
 \,.
\end{equation}
One checks that  the metric is \emph{not flat} at $p$, and hence in a spacetime neighborhood of $p$. (This is a sufficient, but not necessary, condition).

As an example, let us take smooth rotation invariant initial data for $\psi$ at $t=0$ such that $\psi|_{t=0} \equiv 0$, $\partial_t \psi|_{t=0,\, r\le 1} \equiv 0$, and $\partial_t \psi |_{t=0,\, r >1}> 0$. Then  \eqref{30VI22.1} holds at $ t=0$ and $r>1$. This results, by evolution, in a smooth spacetime metric with two spacelike Killing vectors, $\partial_\varphi$ and $\partial_z$, with a spacetime metric which is flat in a spacetime neighborhood of the solid cylinder $\{t=0,\,r<1\}$, but is not flat. 
Unique continuation   of the curvature tensor for real-analytic metrics shows that the spacetime metric  cannot be  real-analytic on a maximal hypersurface which is transverse to some linear combination of the spacelike Killing vectors  and which intersects the cylinder $\{t=0,\,r=1\}.$

\bibliographystyle{amsplain}
\bibliography{MorreyProblem-minimal}

\end{document}